# Manipulating THz Spin Current Dynamics by the Dzyaloshinskii-Moriya Interaction in Antiferromagnetic Hematite

Hongsong Qiu[1†], Tom Sebastian Seifert[2†], Lin Huang[3†], Yongjian Zhou[3], Zdeněk Kašpar[2], Caihong Zhang[1], Jingbo Wu[1], Kebin Fan[1], Qi Zhang[4], Di Wu[5], Tobias Kampfrath[2*], Cheng Song[3*], Biaobing Jin[1*], Jian Chen[1], Peiheng Wu[1]

[1]Research Institute of Superconductor Electronics (RISE), School of Electronic Science and Engineering, Nanjing University, Nanjing, PR China.

[2]Institute of Physics, Freie Universität Berlin, D-14195 Berlin, Germany

[3]Key Laboratory of Advanced Materials (MOE), School of Materials Science and Engineering, Tsinghua University, Beijing, China

[4]Department of Physics, Nanjing University, Nanjing, PR China.

[5]National Laboratory of Solid State Microstructures, Jiangsu Provincial Key Laboratory for Nanotechnology, Collaborative Innovation Center of Advanced Microstructures and Department of Physics, Nanjing University, Nanjing, PR China.

†These authors contributed equally: Hongsong Qiu, Tom Sebastian Seifert, and Lin Huang

*Correspondence should be addressed to Tobias Kampfrath, Cheng Song, and Biaobing Jin: tobias.kampfrath@fu-berlin.de, songcheng@mail.tsinghua.edu.cn, and bbjin@nju.edu.cn.

## Abstract

An important vision of modern magnetic research is to use antiferromagnets as controllable and active ultrafast components in spintronic devices. Hematite is a promising model material in this respect because its pronounced Dzyaloshinskii-Moriya interaction leads to the coexistence of antiferromagnetism and weak ferromagnetism. Here, we use femtosecond laser pulses to drive terahertz spin currents from hematite into an adjacent Pt layer. We find two contributions to the generation of the spin current with distinctly different dynamics: the impulsive stimulated Raman scatting that relies on the antiferromagnetic order and the ultrafast spin Seebeck effect that relies on the net magnetization. The total THz spin current dynamics can thus be manipulated by a medium-strength magnetic field. The controllability of the THz spin current achieved in hematite opens the pathway toward controlling the exact spin current dynamics from ultrafast antiferromagnetic spin sources.

## Introduction

Taming antiferromagnetism is a great challenge for modern magnetism research[1–5]. Though notoriously difficult to manipulate[6–8], antiferromagnets (AFMs) keep fascinating researchers because of the stability against an external magnetic perturbation and the potential for ultrafast operations in the terahertz (THz) frequency range[9–13]. In recent years, miscellaneous experimental strategies, e.g., THz-driven linear[14–17] and nonlinear magnon responses[18,19], THz magnon-phonon coupling[20–23], and THz magnetoelectric coupling[24,25], provided deep insights into the ultrafast response of AFMs. As an emerging phenomenon, AFM spin pumping without the need for a strong external magnetic field is realized by the use of laser-induced THz spin currents[26,27]. It prospectively devises a feasible scheme for practical antiferromagnetic spintronic devices, and more efforts on this topic are urgently required.

Hematite ($\alpha$-Fe$_2$O$_3$) is ubiquitous on earth, and its antiferromagnetic properties have been long studied[28–31]. It belongs to the trigonal crystal system, and the two magnetic sublattices antiferromagnetically align within the basal plane (0001) between the Morin temperature $T_M \approx 260$ K and Néel temperature $T_N \approx 960$ K. The presence of the Dzyaloshinskii-Moriya interaction (DMI), described by the antisymmetric term in the exchange interaction Hamiltonian, gives rise to a small net magnetization $\boldsymbol{M}$ by slightly canting the two spin sublattices[32,33]. A relatively low spin-flop field (<1 T) can align the Néel vector $\boldsymbol{L}$ perpendicular to the external field direction[30]. The response of $\alpha$-Fe$_2$O$_3$ to moderate magnetic fields makes it a widely studied antiferromagnetic material in spintronics[34–36]. Recently, the dc spin pumping by the acoustic resonant mode in $\alpha$-Fe$_2$O$_3$ enhanced by the DMI is reported[37]. However, whether $\alpha$-Fe$_2$O$_3$ can generate THz spin currents upon ultrafast laser excitation remains an open question.

In this paper, we present the coexistence of two mechanisms for the generation of THz spin currents in $\alpha$-Fe$_2$O$_3$. At zero magnetic fields, antiferromagnetic spin pumping by impulsive stimulated Raman scattering process initiates the injection of spin momentum from the $\alpha$-Fe$_2$O$_3$ layer to an adjacent Pt layer. This spin current can be superimposed by a considerable contribution of the ultrafast spin Seebeck effect when applying an external magnetic field as a direct consequence of the DMI. This tunability of the THz spin current polarity achieved in $\alpha$-Fe$_2$O$_3$ by an external magnetic field provides more flexibility for high-speed antiferromagnetic spintronic devices.

## Results

**Experimental Geometry.**

Figure 1**a** illustrates the measurement scheme of the transmission-type THz emission spectroscopy (See Methods). The coordinate system (*xyz*) is defined in the laboratory frame. The linearly or circularly polarized pump laser is incident along the *z*-axis. For linear laser polarization, the polarization direction is denoted as $\theta$. The 20-nm thick (0001)-oriented $\alpha$-Fe$_2$O$_3$ is grown on the Al$_2$O$_3$ substrate and the capping heavy metal (HM) Pt (thickness of 3 nm) is grown *in situ* (See Methods and Supplementary Section 1). The samples are arranged in an external magnetic field $\boldsymbol{B}_{\text{ext}}$ that is perpendicular to the *z*-axis. The angle between the [11$\bar{2}$0] axis and the *x*-axis is referred to as *β*. The laser-induced THz signal from the samples propagates along the *z*-axis and is probed via the linear electro-optic effect in a 1-mm-thick (110)-oriented ZnTe crystal.

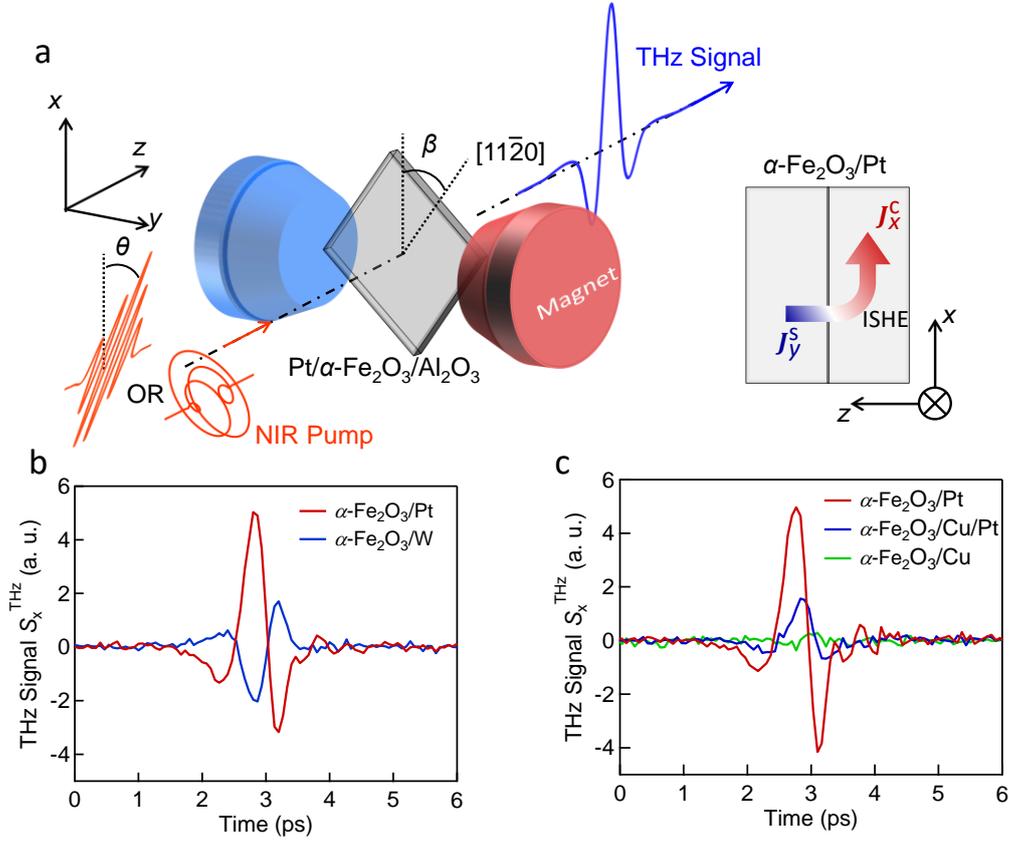

**Fig. 1.** THz signals from the $\alpha$-Fe$_2$O$_3$/HM structures. **a**, Schematic of the transmission-type THz spectroscopy setup. The coordinate (*xyz*) is adapted to the laboratory frame. The sample is placed in the *x-y* plane and the laser is incident along the *z*-direction. The magnetic field $\boldsymbol{B}_{\text{ext}}$ is applied along $\boldsymbol{y}$ (as shown in **a**) or $\boldsymbol{x}$ (not shown). The polarization angle of the pump laser and the rotation angle of the sample are defined as *θ* and *β*, respectively. The sketch of the injection of the spin current $J_y^s$ from $\alpha$-Fe$_2$O$_3$ to the Pt layer is illustrated on the right. The spin current polarized along $\boldsymbol{y}$ is converted to the charge current $J_x^c$ along $\boldsymbol{x}$. **b**, THz signal $S_x^{\text{THz}}$ of the $\alpha$-Fe$_2$O$_3$/Pt sample for $B_{\text{ext}}$ = 0. The polarities of THz signals from $\alpha$-Fe$_2$O$_3$/Pt (red) and $\alpha$-Fe$_2$O$_3$/W (blue) are opposite. **c**, THz signal from the $\alpha$-Fe$_2$O$_3$/Cu/Pt (blue) is weaker than that from $\alpha$-Fe$_2$O$_3$/Pt (red). The THz signal from $\alpha$-Fe$_2$O$_3$/Cu (green) is negligible. $\theta$, $\beta$ = 0 ° in **b** and **c**. a. u., arbitrary units.

The red curve in Fig. 1**b** shows a typical waveform of the THz signal $S_x^{\text{THz}}$ (*x*-component of the THz electric field) from $\alpha$-Fe$_2$O$_3$/Pt obtained with conditions $\theta$ = 0 °, $\beta$ = 0 °, and $B_{\text{ext}}$ = 0. The THz spectrum covers the range from 0 to 3 THz (Supplementary Section 2). $S_x^{\text{THz}}$ is confirmed to be linearly polarized by checking the two orthogonal electric components with two combined wire grid polarizers. The THz signal amplitude has a linear relationship with the laser fluence within the pulse-energy range of interest (Supplementary Section 3).

**THz spin current in the $\alpha$-Fe$_2$O$_3$/Pt bilayer.**

To reveal the origin of $S_x^{\text{THz}}$, we prepared control samples $\alpha$-Fe$_2$O$_3$, $\alpha$-Fe$_2$O$_3$/W, $\alpha$-Fe$_2$O$_3$/Cu, and $\alpha$-Fe$_2$O$_3$/Cu/Pt. In Fig.1**b**, the polarity of $S_x^{\text{THz}}$ from $\alpha$-Fe$_2$O$_3$/Pt is opposite to that from $\alpha$-Fe$_2$O$_3$/W. It agrees well with the fact that the spin Hall angles of Pt and W have opposite signs[38]. A 3-nm-thick Cu interlayer attenuates $S_x^{\text{THz}}$ of $\alpha$-Fe$_2$O$_3$/Cu/Pt to less than half of that of $\alpha$-Fe$_2$O$_3$/Pt (Fig. 1**c**). It is explained as that spin current flows through the Cu layer from the $\alpha$-Fe$_2$O$_3$ (Py) into the Pt layer and undergoes losses during transmission[26,39]. The THz signals from the bare $\alpha$-Fe$_2$O$_3$ film and the $\alpha$-Fe$_2$O$_3$/Cu structure are much smaller, indicating the HM layer is indispensable for a strong THz emission.

The aforementioned standard tests confirm the flow of an ultrafast spin current in Pt and support the following scenario: The spin current $\boldsymbol{J}^s(t)$ is injected from the $\alpha$-Fe$_2$O$_3$ layer into the Pt layer, as illustrated on the right of Fig. 1**a**; $\boldsymbol{J}^s(t)$ is converted into an in-plane charge current $\boldsymbol{J}^c(t)$ in the Pt layer because of the inverse spin Hall effect[40]. The transient $\boldsymbol{J}^c(t)$ emits a THz wave $S_x^{\text{THz}}$ into free space. The THz spin current polarized along $\boldsymbol{y}$ ($J_y^s$) can be retrieved from $S_x^{\text{THz}}$ by taking advantage of the impulse response function of the THz emission setup (see Methods). In the following, we focus on $J_y^s$.

**Opto-magnetic origin of the THz spin current at zero magnetic fields.**

Generally, the ultrafast spin injection can be realized by the incoherent driving forces: pyrospintronic effect (PSE)[41–43] and ultrafast spin Seebeck effect (SSE)[39,42,44] and the coherent driving forces: impulsive stimulated Raman scattering (ISRS)[26,27] and strain wave[27]. The incoherent driving forces PSE and SSE are heating-induced spin voltage and temperature gradient at the interface AFM/HM, respectively. They require non-zero preexisting net magnetization in the spin source. In contrast, the coherent driving forces induce impulsive magnetization in the AFM layer and pump the spin current into the HM layer, wherein preexisting net magnetization is not mandatory.

At room temperature, the as-grown $\alpha$-Fe$_2$O$_3$ film is expected to contain magnetic domains orienting randomly along all easy axes ($\langle 1\bar{1}00 \rangle$ axes). The grain size of each single spin domain $\sigma$ is of the order of 1 μm[45]. Therefore, within the area of the laser spot (diameter of ~ 3 mm), the net magnetization $\sum_\sigma \boldsymbol{M}$ averages to approximately zero. We, thus, ascribe the generation of $J_y^s$ to ultrafast spin pumping launched by the laser-induced transient magnetization $\Delta \boldsymbol{M}(t)$ [26,27,46,47]. The dynamic $\Delta \boldsymbol{M}(t)$ originates from an effective magnetic field that is induced by the optical field through an ISRS process, as observed in AFMs previously[48–50].

An off-resonant response is expected to sensitively depend on the pump polarization and is, thus, tested by studying the impact of the pump polarization on $J_y^s$. As shown in Fig. 2**a**, at a sample azimuth of $\beta = 180°$, a linearly polarized pump beam is more efficient in generating $J_y^s$ than a circularly polarized one. In addition, the polarity of the traces is reversed for $\theta = 0°$ (red) and $\theta = 90°$ (blue) in the case of linearly polarized excitation. In contrast, at $\beta = 270°$, a circularly rather than a linearly polarized pump facilitates the generation of $J_y^s$ (Fig. 2**b**). The reversed sign of the current traces is compelling evidence for the pump-helicity dependence.

Note that both Fig. 2**a** and 2**b** indicate a relatively small $J_y^s$ component with minor dependence on the pump polarization. We ascribe this polarization-insensitive contribution to the strain wave generated in the heating process of the Pt layer as observed previously[27,30,51,52]. The slightly different dynamics for the polarization-insensitive contributions at different $\beta$ might indicate an anisotropic magnetoelastic coupling and requires further investigation. Importantly, the strong dependence of the dominant $J_y^s$ contributions on the laser polarization state is a major indication for the ISRS process[49,53].

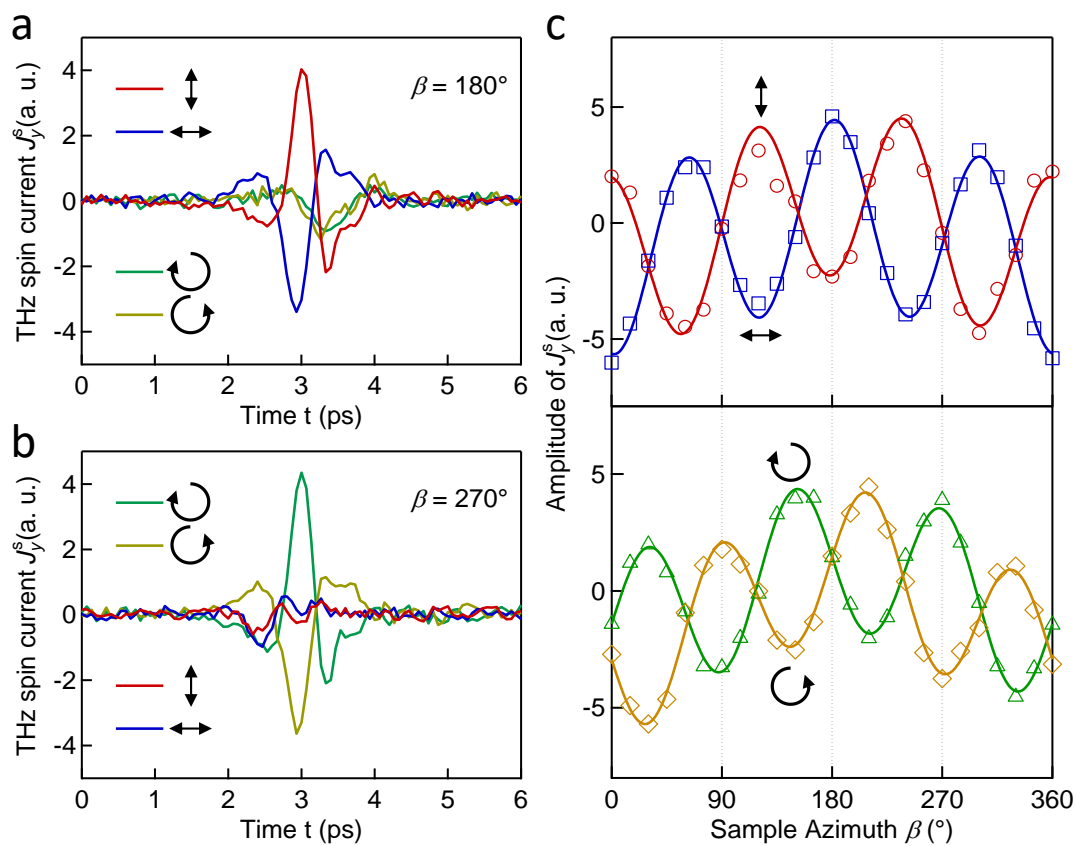

**Fig. 2.** Impact of the state of the laser polarization at $\sum_\sigma M = 0$. The traces of $J_y^s$ for various laser polarization states at **a**, $\beta =180°$ and **b**, $\beta =270°$. **c**, The amplitude of $J_y^s$ varies as a cosine function when using the linear laser (top) and as a sine function when using the circular laser (bottom). The marks are measured results and the solid curves are fit. Laser polarization state: Red, along *x*; Blue, along *y*; Green, left; Yellow, right.

In principle, the off-resonant $J_y^s$ component can be launched through ISRS that depend on the magnetic order parameter in first order (inverse Faraday effect (IFE)) or second order (inverse Cotton-Mouton effect (ICME))[49]. Interestingly, Fig. 2**a** and 2**b** imply that ICME and IFE dominate for complementary orthogonal sample azimuths. The dependence of the amplitude of $J_y^s$ on $\beta$ is compared in detail for the cases of ICME and IFE (Fig. 2**c**). The threefold period in all curves is consistent with the trigonal symmetry of the (0001)-oriented $\alpha$-Fe$_2$O$_3$ film. Notably, $J_y^s$ associated with ICME varies as a cosine function, while that associated with IFE varies as a sine function. The reason is that the polarization of the spin current generated by the linearly polarized pump is mainly determined by **M**, while that generated by the circularly polarized pump is mainly determined by **L** (see Methods).

In summary, we can denote the THz spin current observed at zero magnetic fields as $J_y^{s,\mathrm{ISRS}}$, with a superscript indicating the ISRS origin. It can be phenomenologically described as the temporal convolution of the opto-magnetic coefficients for the ISRS and the laser fluence[26]

$$J_y^{s,\mathrm{ISRS}}(t) = \sum_\sigma \left( \chi_{yii}^{\mathrm{lin}}(M_{x'}) * E_i E_i^* + \chi_y^{\mathrm{cir}}(L_{y'}) * \left( E_x E_y^* - E_x^* E_y \right) \right)(t), \quad (1)$$

with **E** being the electric field of the pump laser pulse. The opto-magnetic coefficients $\chi_{yii}^{\mathrm{lin}}$ ($i = x$ or $y$) and $\chi_y^{\mathrm{cir}}$ are obtained by transforming the local coefficients in the spin coordinate $(x'y'z')$ to the lab coordinate ($xyz$). The summation convention for repeated indices is applied on the $\chi_{yii}^{\mathrm{lin}}$ term. The total spin current is obtained by summing the contribution from each magnetic domain $\sigma$. The fit (solid curves in Fig. 2**c**) based on Eq. (1) has a quantitative agreement with the experimental data (marks).

**Manipulation of the THz spin current by an external magnetic field.**

A non-zero net magnetization $\sum_\sigma M \neq 0$ appears when an external magnetic field is applied to the $\alpha$-Fe$_2$O$_3$ film. We study the influence of $\sum_\sigma M \neq 0$ on $J_y^s$ by scanning $B_\mathrm{ext} \parallel y$ in the range from -1 T to 1 T. As seen in Fig. 3**a**, $J_y^s$ exhibits a hysteretic feature in both measurements conducted with linearly (red) and circularly (blue) polarized laser pulses, which is in stark contrast to the linear magnetic response of KCoF/Pt and KNiF/Pt structures[54]. The solid curves are a sigmoid fit to the experimental data, yielding a coercivity field lower than 0.15 T (Supplementary Section 4). The hysteretic response of $J_y^s$ is highly consistent with magnetic-moment measurements via a superconducting quantum interface device (SQUID) at 300 K

(Supplementary Section 1). We, thus, conclude that there is an additional contribution to $J_y^s$ by the nonzero net magnetization $\sum_\sigma M \neq 0$.

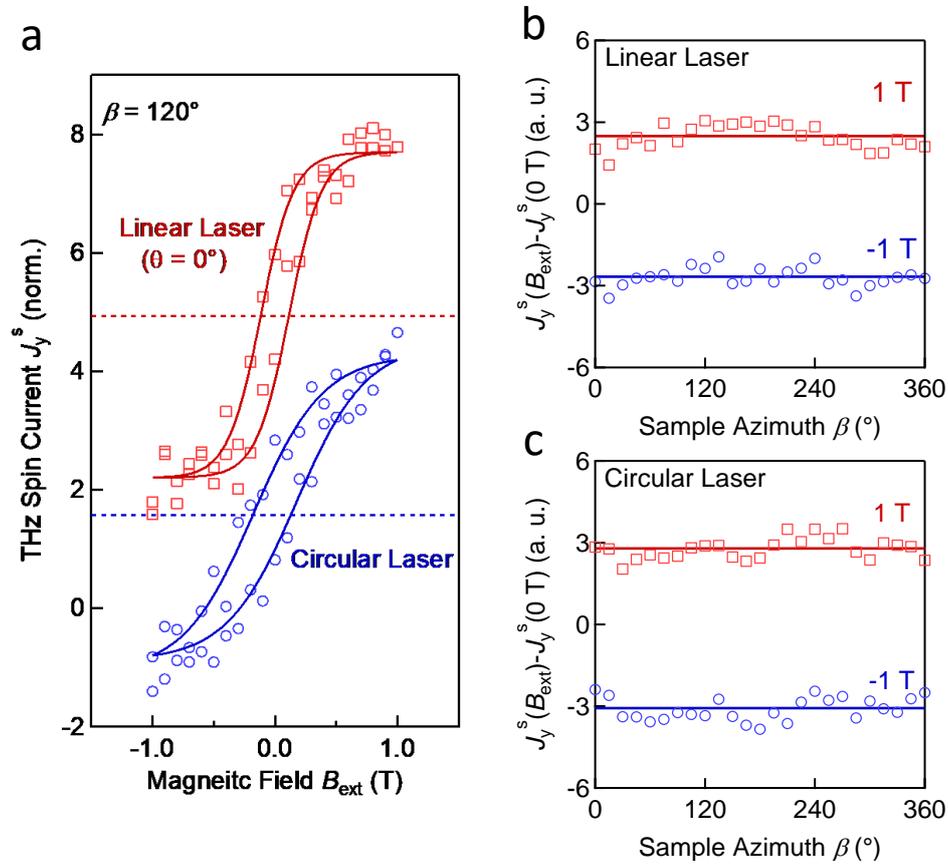

Fig. 3. Manipulation of $J_y^s$ by $B_{ext}$. **a**, Amplitude of $J_y^s$ vs $\boldsymbol{B}_{ext} \parallel \boldsymbol{y}$ for linearly (red) and circularly (blue) polarized pump pulses and $\beta=120°$. The solid curves are sigmoid fits. **b**, The additional contribution to $J_y^s$ while applying $|\boldsymbol{B}_{ext}| = 1\,\text{T}$ is extracted in the way of $J_y^s(H_{ext}) - J_y^s(0\,\text{T})$. A linear pump polarization ($\theta = 0$) is used in the measurement. **c**, Analogous to panel **b** for circular pump polarization. The solid lines indicate the average values of $J_y^s$ when $B_{ext}$ is 1 T (red) or -1 T (blue). Note that the vertical offset of all data is as measured.

Note that two hysteresis loops in Fig. 3**a** exhibit vertical offsets. The hysteresis loops for various $\beta$ are shown in Supplementary Section 4. The horizontal dashed line in each loop shows the average value during the fitting, which is close to the amplitude of $J_y^{s,\text{ISRS}}$ that is obtained for $\sum_\sigma M = 0$ (Fig. 2**c**). Thus, the additional contribution to $J_y^s$ by the nonzero net magnetization $\sum_\sigma \boldsymbol{M} \neq 0$ is superimposed on $J_y^{s,\text{ISRS}}$ as a constant part that does not change with pump polarization or sample azimuth.

As shown in Fig. 3**a**, the polarization state (linear or circular) of the pump laser does not affect the amplitude of the additional spin-current contribution for $\sum_\sigma \boldsymbol{M} \neq 0$. For better comparison, we extract the additional contribution by calculating the difference $J_y^s(H_{\text{ext}}) - J_y^s(0\text{ T})$ for various $\beta$ when $\boldsymbol{B}_{\text{ext}}$ is kept at 1 T or -1 T. Both for using linear (Fig. 3**b**) and circular (Fig. 3**c**) laser, $J_y^s$ contains a large $\beta$-independent offset superimposed by a minor fluctuation. Similar results can also be observed by studying the dependence of $J_y^s$ on $\theta$ (Supplementary Section 5). The minor fluctuation is probably attributed to the small changes in the opto-magnetic coefficient $\chi_{ijkl}M_k$ for the ISRS under the influence of $\boldsymbol{B}_{\text{ext}}$[53]. The sizeable $\beta$-independent part, as denoted by the solid horizontal line in Fig. 3**b** and 3**c**, is odd in $\boldsymbol{B}_{\text{ext}} \parallel \boldsymbol{y}$ and not impacted by the state of the pump pulse polarization. This observation indicates the emergence of a spin-current contribution for $\sum_\sigma \boldsymbol{M} \neq 0$ in addition to the ultrafast off-resonant contribution that dominates at $\sum_\sigma \boldsymbol{M} = 0$.

**Spin-caloritronic contribution to the THz spin current at a finite external magnetic field.**

Coherent and incoherent driving forces exhibit different temporal evolution of the ultrafast spin current. Therefore, the time scale of the ultrafast spin current evolution can be used as the hallmark to clarify its origin[42]. To capture the ultrafast evolution of $J_y^s$, we conduct a measurement in the THz emission setup based on a 15-fs Ti:sapphire laser oscillator (See Methods).

As shown in Fig. 4**a**, the odd (red) and even (blue) components of the THz signal $S_x^{\text{THz}}$ in the magnetic field are extracted by taking the difference $S_x^{\text{THz}}(B_{\text{ext}}) - S_x^{\text{THz}}(-B_{\text{ext}})$ and the sum $S_x^{\text{THz}}(B_{\text{ext}}) + S_x^{\text{THz}}(-B_{\text{ext}})$, respectively ($B_{\text{ext}} = 0.4$ T). The odd THz signal of a fully metallic Fe/Pt thin-film structure (green) is shown for comparison. The fast even component has an impulsive feature, basically following the pump-pulse intensity envelope superimposed by a fast oscillatory signal. The high-frequency oscillation is attributed to the excitation of phonon modes in $\alpha$-Fe$_2$O$_3$ within the frequency range from 10 to 20 THz[48,55,56]. Its detailed origin requires further studies.

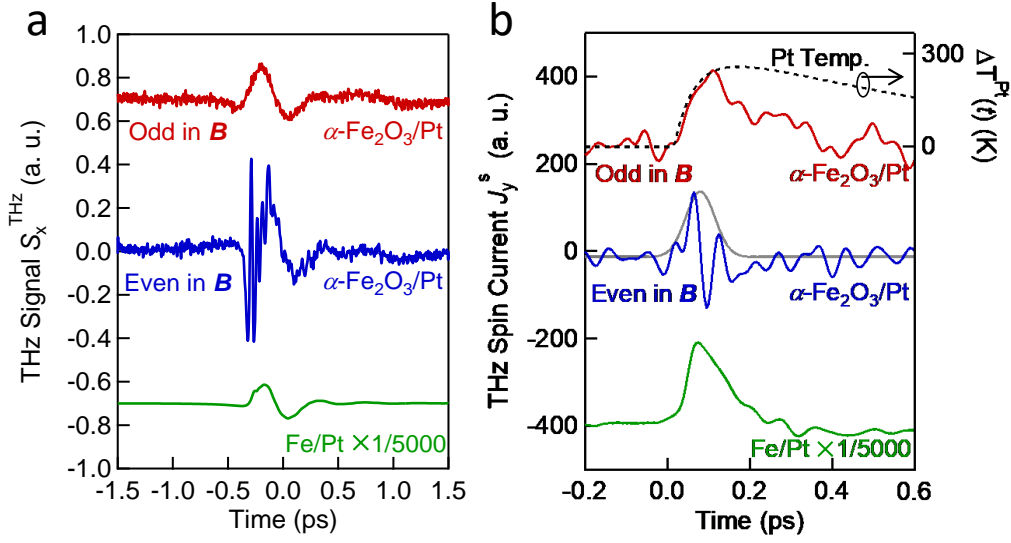

Fig. 4. Temporal evolution of $J_y^s$ in $\alpha$-Fe$_2$O$_3$/Pt and Fe/Pt. **a**, Components of the THz signal $S_x^{THz}$ that is odd (red) and even (blue) in the external magnetic field $B_{ext} = 0.4$ T. The green curve shows the evolution of $J_y^s$ in Fe/Pt as a comparison. **b**, $J_y^s$ is retrieved from **a**. Compared with in Fe/Pt (green), the $J_y^s$ component odd in the external magnetic field (red) in $\alpha$-Fe$_2$O$_3$/Pt slowly evolves with a rise time >100 fs, whereas the fast even component (blue) has an impulsive feature, basically following the pump-pulse intensity envelope superimposed by a fast oscillatory signal. The dotted curve is the calculated generalized electronic temperature in Pt following laser excitation (Adapted from [39]). The grey curve is a Gaussian fit with a full width at half maximum of ~50 fs. Note that the curves in **b** were smoothed with a Gaussian function with a 1/e width of 26 fs for better visibility.

The waveforms of spin current $J_y^s$ are retrieved from Fig. 4**a** and correspondingly shown in Fig. 4**b**. The high-frequency oscillation is removed in Fig. 4**b** by smoothening the curve with a Gaussian function with a 1/e-width of 26 fs. The even component of $J_y^s$ (blue) approximately equals $J_y^{s,ISRS}$ and is launched by the ISRS dynamics. In principle, the even component could include nonmagnetic signal contributions, which we, however, assume to be minor due to the absence of any detectable THz emission in the α-Fe$_2$O$_3$/Cu control sample (Fig. 1c). As a reference, we compare our extracted spin currents to that of a fully metallic Fe/Pt thin-film structure (green), which has a relatively rapid rise and decay. The driving force in Fe/Pt is the PSE, and the time scale of its relaxation is mostly determined by the electron-spin equilibration time in the ferromagnetic metal[43]. In contrast, the temporal evolution of the odd component of $J_y^s$ (red) in the α-Fe$_2$O$_3$/Pt structure exhibits a slower rise time of more than 100 fs and an even slower decay. These markedly different time scales strongly suggest driving forces different from the ISRS process[27] and the PSE and rather indicate that the odd component of $J_y^s$ is dominated by the ultrafast SSE.

The temperature gradient across the interface of the magnetic and the paramagnetic layer is necessary for the ultrafast SSE. In the case of a magnetic insulator, laser-excited hot electrons in the metal layer (Pt) get spin-polarized upon scattering off the interface towards the magnetic insulator[57]. In α-Fe$_2$O$_3$/Pt, the electronic temperature of the Pt layer increases by $\Delta T^{Pt}(t)$ upon laser excitation[58,59], while that of the α-Fe$_2$O$_3$ layer remains unaffected because of the weak absorption[60]. Therefore, the ultrafast-SSE spin current injected from the α-Fe$_2$O$_3$ to the Pt layer can be expressed as the temporal convolution $J_y^{s,\Delta T}(t) = (\kappa_y^{Pt} * \Delta T^{Pt})(t)$, with $\kappa_y^{Pt}(t)$ being the response function that relates the spin current in the Pt layer to an ultrashort delta-like temperature increase of the Pt electrons. This response function is proportional to the convolution of the spin susceptibilities of the Pt and the α-Fe$_2$O$_3$ layer.

**Quasi-instantaneous fit to the SSE spin current**

While $\Delta T^{Pt}(t)$ has an ultrafast rise and a decay time on the order of 100 fs, $\kappa_y^{Pt}(t)$ has a much shorter duration and is dominated by the spin-spin correlation time in Pt, which is of the order of a few femtoseconds only[39]. As a result, $\kappa_y^{Pt}$ acts like a δ-like

function, and $J_y^{s,\Delta T}$ follows $\Delta T^{Pt}(t)$ quasi-instantaneously:

$$J_y^{s,\Delta T} = \eta\left(\sum_\sigma M_y\right)\Delta T^{Pt}(t). \qquad (2)$$

The SSE coefficient $\eta$ is odd in the net magnetization and, thus, scales linearly with $\sum_\sigma M_y$ to lowest order. As excited by the ultrafast laser pulses, highly energetic electrons in Pt are generated. Subsequently, the electrons thermalize via secondary scattering cascades to form more carriers above the Fermi energy yet with individually less energy. The calculated temperature evolution $\Delta T^{Pt}(t)$ for the electrons in Pt is shown by the dashed curve in Fig. 4**b** (See Methods). Based on these considerations, we can understand the apparently good agreement between $\Delta T^{Pt}(t)$ and $J_y^{s,\Delta T}$.

One should, however, note that, in YIG/Pt, γ-Fe$_2$O$_3$/Pt, and Fe$_3$O$_4$/Pt structures[42], the evolution of the spin current (rise time >200 fs) lags behind that of the calculated temperature by ~100 fs. We speculate that these discrepancies can arise from either altered thermalization dynamics in Pt layers grown under different conditions on different magnets (electron scattering times might alter significantly) or a direct excitation of carriers in α-Fe$_2$O$_3$ due to its relatively small band gap of ~2.1 eV in comparison to the pump photon energy of 1.5 eV[61]. The latter scenario might lead to a spin-voltage-like driving force that entails a contribution with faster dynamics than the typical SSE current measured in previous experiments. Indeed, we find that the measured spin current odd in magnetization in α-Fe$_2$O$_3$/Pt can be reproduced by a linear combination of the spin current driven by the ultrafast spin voltage in Fe/Pt and the ultrafast Seebeck current in a YIG/Pt sample (Supplementary Section 6).

## Discussion

Our analysis indicates that the total THz spin current can be rewritten as $J_y^s = J_y^{s,ISRS} + J_y^{s,\Delta T}$. The insensitivity of the laser-induced $\Delta T^{Pt}(t)$ to the laser polarization[39,62] is in line with the $\beta$-independent contribution shown in Fig. 3**a** and 3**b**. Besides, according to Eq. 2, $\sum \boldsymbol{M} \parallel \boldsymbol{x}$ does not contribute to $J_y^s$ via the ultrafast SSE.

As a test, the dependence of $J_y^s$ on $\beta$ is measured when $\boldsymbol{B}_{\text{ext}} \parallel \boldsymbol{x}$ (1 T) is applied (Supplementary Section 7). The absence of the $\beta$-independent modulation of $J_y^s$ by $\boldsymbol{B}_{\text{ext}} \parallel \boldsymbol{x}$ fully agrees with our interpretation including the ultrafast SSE.

Hematite is classified as a g-wave altermagnet with four nodal surfaces in the Brillouin zone[63,64]. Because g-wave altermagnets have a spin-independent averaged electrical conductivity, the contribution to the giant magnetoresistance from the spin-dependent averaged conductivities is absent[65,66]. Analogously, we expect no altermagnetism-related contributions to the SSE, assuming that the thermal spin transport follows the same symmetry rules as the electrical spin transport.

We can also exclude other signal contributions that are insensitive to the pump polarization. (i) The optical perturbation of the exchange interaction in $\alpha$-Fe$_2$O$_3$ via a virtual electric-dipole transition from $2p$ orbitals of $O^{2-}$ to $3d$ orbitals of $Fe^{3+}$ in the iron-oxygen cluster[48]. This mechanism is expected to have much faster dynamics than $\Delta T^{\text{Pt}}(t)$. (ii) A laser-induced rapid change of the magnetocrystalline anisotropy can give rise to a spin reorientation in iron oxides by transiently heating above the Morin temperature[9]. For bulk $\alpha$-Fe$_2$O$_3$, the Morin temperature is around 260 K and it considerably decreases for thin films[67,68]. Our measurements were conducted at room temperature (~300 K), which is much higher than the actual Morin temperature. As a result, there is no laser-induced spin reorientation in our experiments.

The even component of $J_y^s$ (blue) obtained by the sum $J_y^s(0.4\,\text{T}) + J_y^s(-0.4\,\text{T})$ shows an impulsive response, that is, it follows the intensity envelope of the pump pulse. The grey curve in Fig. 4**b** is a Gaussian fit with a full width at half maximum of ~50 fs, which combines the pump pulse duration (15 fs) and the bandwidth of the current extraction procedure (1/30 THz= 33 fs) as well as the intrinsic time scale of the ISRS. The latter was calculated to have dynamics of the order of 30 fs[69], which agrees well with our findings.

As a summary, we comprehensively studied the mechanisms for the generation of the THz spin current $J_y^s$ in the $\alpha$-Fe$_2$O$_3$/Pt structure. At $\sum_\sigma \boldsymbol{M} = 0$, the ultrafast spin pumping is facilitated by the ISRS process and dominates the generation of $J_y^s$. ICME and IFE predominate in orthogonal directions. The ultrafast spin pumping acts on a time scale of a few tens of femtoseconds. The ultimate time scale may be still faster because of the finite laser pulse duration. At $\sum_\sigma M \neq 0$, the ultrafast SSE additionally contributes to the generation of $J_y^s$, which evolves slowly with respect to that generated via ultrafast spin pumping.

Multiple spin injection mechanisms typically coexist in magnetic systems. The laser-excited electrons carrying spin angular momentum in metallic systems can inject a spin current driven by the spin voltage. It is overwhelmingly predominant in ferromagnetic metals and largely outperforms other mechanisms. Therefore, the insulating antiferromagnetic $\alpha$-$Fe_2O_3$ provides a platform to observe the ultrafast coherent spin pumping and the ultrafast SSE simultaneously. In contrast, in NiO/Pt structures, there is no obvious THz spin current driven by the ultrafast SSE because of the absence of a net magnetization due to a lack of DMI (Supplementary Section 8). From an applied viewpoint, the spin current related to the Néel vector and the net magnetization can be distinguished by checking the dependence on the laser polarization, which demonstrates a new technical scheme for detecting the detailed spin texture in AFM even on ultrafast time scales.

The presence of DMI causes a weak spontaneous magnetization in $\alpha$-$Fe_2O_3$. Thus, the magnetic state of $\alpha$-$Fe_2O_3$ can be conveniently manipulated by a medium-strength magnetic field (less than 0.4 T in our measurements). Thereby, a spin current odd in the net magnetization, in addition to the ISRS processes, can be driven through the ultrafast SSE. More importantly, the adjustable net magnetization opens an exciting pathway toward the control of the exact spin current dynamics from ultrafast AFM spin sources.

## Methods

**Measurement.** All measurements except that shown in Fig. 4 are conducted with a THz-emission setup based on a Ti:sapphire amplified laser. The laser provides pulses with 100-fs duration, 1-kHz repetition rate, and 800-nm central wavelength. The pump beam was loosely focused on the sample surface with a spot diameter of around 3 mm. The laser fluence was around 1.2 mJ/cm$^2$.

In Fig. 4, the evolution of the spin current is measured with a 15-fs Ti:sapphire laser oscillator (center wavelength 800 nm, pulse energy 2.5 nJ, and repetition rate 80 MHz). The duration of the laser pulses is compressed by using a pair of wedged prisms and a pair of chirped mirrors. The diameter of the pump beam at the sample surface is ~20 μm. The resulting absorbed fluence was 0.12 mJ/cm$^2$.

The magnetic field ($|\boldsymbol{B}_{\text{ext}}| \leq 1$ T) was applied by a d. c. electromagnet. The magnetic field was monitored with a Gauss meter. The divergent THz signal was collected by a parabolic mirror with a reflected focal length (RFL) of 101.6 mm. Another parabolic mirror (RFL = 50.8 mm) focused the collimated THz signal onto the 1-mm-thick (110)-oriented ZnTe crystal with a spot diameter of around 300 μm. The ellipticity modulation of the probe beam that was tightly focused on the center of the THz spot was captured by a pair of balanced photodetectors for the acquisition of the THz signal. All the measurements were carried out at room temperature (~300 K) and in the dry air or N$_2$ atmosphere.

**Sample Fabrication.** The $\alpha$-Fe$_2$O$_3$/Pt samples were grown in a high vacuum magnetron sputtering chamber with a base vacuum of $5 \times 10^{-5}$ Pa. The thickness is 20 nm for the $\alpha$-Fe$_2$O$_3$ layer and 3 nm for the Pt layer. The $\alpha$-Fe$_2$O$_3$ films were grown on (0001)-oriented Al$_2$O$_3$ substrates under an atmosphere (Ar:O$_2$ = 10:1) with the temperature of substrates at 500 °C. The Pt layer was then *in-situ* capped on the $\alpha$-Fe$_2$O$_3$ layer at room temperature. The qualities of films were characterized by X-ray diffraction. All samples in the main text were fabricated on the (0001)-oriented Al$_2$O$_3$ substrates.

**Macroscopic Theory for Ultrafast Spin Pumping.** The Hamiltonian describing the interaction between the laser and the medium is a function of the dielectric tensor $\varepsilon_{ij}$. The laser pulse modulates $\varepsilon_{ij}$, acting as effective fields $\boldsymbol{H}_{\text{eff}}^{M}$ and $\boldsymbol{H}_{\text{eff}}^{L}$, which are expressed as the partial derivative of the interaction Hamiltonian to $\boldsymbol{M}$ and $\boldsymbol{L}$, respectively[70]. Therefore, the ISRS process induces an impulsive magnetization $\Delta \boldsymbol{M}(t)$ in the $\alpha$-Fe$_2$O$_3$ layer[71] and gives rise to spin pumping on a time scale of sub-picoseconds.

The spin pumping in $\alpha$-Fe$_2$O$_3$/Pt is described as[37]

$$\boldsymbol{J}^s = \hbar g_{\text{mix}} \boldsymbol{M} \times \frac{\partial \boldsymbol{M}}{\partial t}, \tag{S6}$$

in which $\hbar$ is reduced Planck constant and $g_{\text{mix}}$ is the interfacial spin-mixing conductance. Here, we use the Landau-Lifshitz-Gilbert equations to describe the dynamics of $\boldsymbol{M}$ and $\boldsymbol{L}$[70]

$$\frac{\partial \boldsymbol{M}}{\partial t} = -\gamma \{\boldsymbol{M} \times \boldsymbol{H}^M_{\text{eff}} + \boldsymbol{L} \times \boldsymbol{H}^L_{\text{eff}}\}, \tag{S5}$$

where $\gamma$ is the gyromagnetic ratio. Combining Eq. S5 and S6, the spin pumping can be written as

$$\boldsymbol{J}^s = -\hbar g_{\text{mix}} \gamma [\boldsymbol{M}(\boldsymbol{M} \cdot \boldsymbol{H}^M_{\text{eff}}) - \boldsymbol{H}^M_{\text{eff}}(\boldsymbol{M} \cdot \boldsymbol{M}) + \boldsymbol{L}(\boldsymbol{M} \cdot \boldsymbol{H}^L_{\text{eff}})] \tag{S7}$$

In our experimental geometry, only the spin current polarized in the plane can be detected. We focus on a single spin domain first. Therefore, we write the spin pumping in the form of

$$\begin{bmatrix} J^s_x \\ J^s_y \end{bmatrix} = -\hbar g_{\text{mix}} \gamma \begin{bmatrix} M_x \boldsymbol{M} \cdot \boldsymbol{H}^M_{\text{eff}} \\ L_y \boldsymbol{M} \cdot \boldsymbol{H}^L_{\text{eff}} + o \end{bmatrix}, \tag{S8}$$

with a minor correction term written as $o = M_y \boldsymbol{M} \cdot \boldsymbol{H}^M_{\text{eff}}$. The value of the $o$-term is related to the coefficient $k_{xyz}$ and tends to be very small, which is experimentally confirmed in Fig. 2**c** of the main text. The *x*-component of the spin current is dominated by $\boldsymbol{H}^M_{\text{eff}}$ while the *y*-component is dominated by $\boldsymbol{H}^L_{\text{eff}}$.

According to Eq. S5 and S6, $\boldsymbol{H}^M_{\text{eff}}$ and $\boldsymbol{H}^L_{\text{eff}}$ are the effective fields generated via IFE and ICME, respectively. As a result, the two mechanisms predominate in orthogonal directions. The striking agreement between the measurement and the theory verifies that the origin of the spin pumping is directly related to the ISRS.

There are three easy axes in the basal plane of hematite. The contribution of all spin domains can be calculated by transforming the tensors in each spin frame into that of the laboratory frame[72]. The final form of the formula can be expressed as the linear superposition of the $\cos(3\theta)$ and $\cos(\theta)$ function[49,72], in which all the parameters are subsumed into the coefficients in front of the trigonometric functions.

The THz spin current originating from the ISRS process can be essentially described as Eq. 1 in the main text. Note that the opto-magnetic coefficients $\chi_{yii}^{\text{lin}}$ ($i = x$ or $y$) or $\chi_y^{\text{cir}}$ is not only determined by $M_{x'}$ or $L_{y'}$ (Supplementary Section 9). The simplified form is adopted in the main text to denote that ICME and IFE predominate in orthogonal directions.

**Extraction of the THz Current.** The current extraction for the $\alpha$-Fe$_2$O$_3$ sample was done by using a reference THz signal from a fully metallic Fe/Pt sample, for which the current was known from the previous studies[42]. We then applied a matrix inversion procedure to deconvolute the detected electro-optic signals by the setup response function in the time domain[39].

## Author Contributions

T. K., C. S., and B. J. conceived the idea and instructed this work. L. H. and Y. Z. fabricated and characterized the hematite samples. H. Q. and C. Z. built the terahertz emission setup based on the 100-fs ultrafast laser system and performed the terahertz experiments. T. S. S. and Z. K. performed the measurements with the terahertz setup constructed in a 15-fs ultrafast laser system. J. W., K. F., Q. Z., D. W., J. C., and P. W. provided very revealing comments on the physical mechanism of the ultrafast interaction between light and matter. H. Q., T. S. S., and L. H. analyzed experimental data and wrote the manuscript with contributions from all the authors.